# Dynamic Metasurface Antenna Enabled Real-Time 60 GHz Wireless Video Streaming: A System-Level Demonstration with USRPs and mmWave Front-Ends

Abdul Jabbar*

*Wolfson School of Mechanical, Electrical and Manufacturing Engineering, Loughborough University, UK

***Abstract*—This paper presents a real-time 60 GHz wireless video transmission system enabled by a millimeter-wave (mmWave) Dynamic Metasurface Antenna (DMA) and offers a 60 GHz transceiver solution. The end-to-end communication system is established using off-the-shelf mmWave up/down-conversion modules (EK1HMC6350, HMC6300, and HMC6301) with a connectorized DMA-based transmit front-end and universal software radio peripheral (USRP) for digital baseband processing. The proposed framework eliminates the need for costly high-frequency signal generators and spectrum analyzers, thus simplifying implementation, validation, and reducing operational costs. A real-time HD video stream at 720p resolution (1280 × 720) is transmitted and successfully retrieved over a wireless link using QPSK modulation, achieving a data rate of about 3 Mbps with high reliability. The demonstrated platform can support various 60 GHz standard physical-layer protocols, such as IEEE 802.11ad/ay and 802.15.3c. This work highlights the potential of programmable metasurface antennas as a compact, energy-efficient alternative to legacy phased arrays for next-generation 6G and beyond mmWave/sub-THz wireless communication systems, small-cell outdoor backhaul links, and RF sniffing and channel sounding.**

**Index Terms—4-QAM, 60 GHz, metasurface antenna, QPSK, SDR, transceiver, USRP, wireless communication.**

## I. INTRODUCTION

The rapid growth of data-intensive applications, including high-definition (HD) video streaming, extended reality, and wireless backhaul, is driving the demand for high-capacity wireless links in the millimeter-wave (mmWave) and sub-terahertz (THz) spectrum. The unlicensed 60 GHz industrial, scientific, and medical (ISM) band is particularly attractive due to its multi-gigahertz bandwidth, enabling gigabit-per-second data rates for real-time indoor communication systems such as industrial wireless networks, as well as outdoor small cell backhaul use cases [1] . However, the deployment of reliable mmWave links remains challenging due to severe free-space path loss, susceptibility to blockage, and quasi-optical propagation characteristics [2], [3]. These challenges demand highly directive antennas and carefully engineered end-to-end radio-frequency (RF) systems [4].

Often, mmWave communication system validation relies on horn antennas due to their simplicity, ease of availability, and well-defined radiation characteristics. Nevertheless, their relatively wide beamwidth (typically >15°) and static beam configuration limit link directivity, reducing spatial selectivity and flexibility in practical deployments. Phased array antennas offer beamforming capabilities to address these limitations. However, they require complex RF chains, high power consumption, and costly hardware, which hinder scalability for compact and low-cost mmWave communication systems [5].

Dynamic Metasurface Antennas (DMAs) have recently emerged as a promising alternative to conventional phased array-based beamforming solutions [6]–[8]. This cutting-edge antenna technology enables software-defined wavefront engineering using subwavelength meta-elements, significantly reducing hardware complexity and eliminating the need for power-hungry phase shifters [9]. Despite their potential, experimental demonstrations of DMA-enabled communication systems at mmWave frequencies, particularly for real-time applications, remain limited. Nevertheless, some customized modular 60 GHz front-end designs for RF channel sounding have been proposed [10], [11]. Besides, some open-source customized [12], [13] front-ends as well as SDR-based [14] platforms have also been demonstrated for 60 GHz mmWave systems. Furthermore, programmable reflective metasurfaces in the sub-6 GHz band have also been designed and employed for real-time wireless video and image transmission using SDRs [15], [16]. However, these metasurfaces must be excited by an additional external horn antenna, which makes the overall wireless system quite bulky. Therefore, the direct integration of mmWave programmable metasurface antennas into compact, end-to-end communication systems using low-cost commercial-off-the-shelf (COTS) front-ends and flexible baseband processing remains a significant gap in current research and demands practical exploration [17].

The 60 GHz band offers significant advantages for next-generation high-data-rate industrial wireless applications, and provides reduced installation and maintenance costs, improved flexibility, and inherent interference isolation from crowded sub-6 GHz ISM bands [18]. Its large available bandwidth (up to 14 GHz (57-71 GHz)) allows substantially higher data rates, while its quasi-optical propagation can be exploited for spatially confined and secure communication links [19]–[21].
However, these same characteristics necessitate precise beam control and system-level integration, underscoring the importance of experimental validation with complete end-to-end communication setups rather than isolated component-level analysis.

In this work, a complete system-level demonstration of real-time 60 GHz wireless video transmission is presented, and an end-to-end transceiver solution is elucidated. Instead of relying on costly mmWave equipment such as signal generators and spectrum analyzers, a fully integrated end-to-end testbed using COTS 60 GHz up/down-conversion analog RF front-end modules for high-frequency generation and reception is developed. Interface with Ettus Universal Software Radio Peripheral (USRP)-based software-defined radio (SDR) systems enables flexible digital baseband processing and real-time system operation. The proposed system employs a connectorized DMA [6] with a directive fan-beam radiation pattern as the transmit (Tx) antenna for over-the-air (OTA) wireless transmission and a V-band horn antenna as the receiving (Rx) antenna.

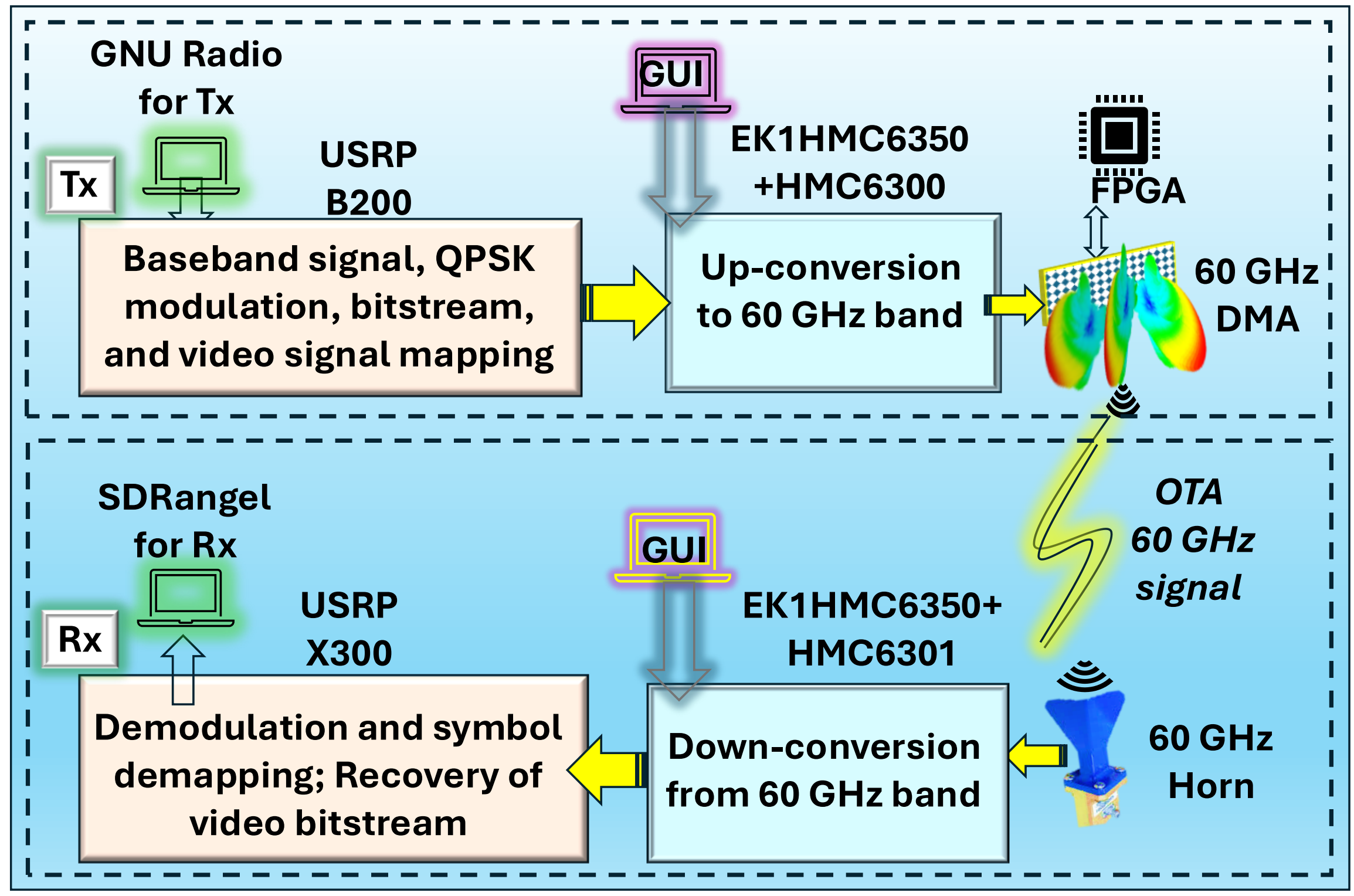

Fig. 1. Block diagram illustrating experimental setup.

A live HD video stream is transmitted using Quadrature Phase Shift Keying (QPSK), equivalently 4-QAM, OTA through a 60 GHz RF channel and successfully recovered at a distant receiver in real time, demonstrating stable link establishment and reliable performance. This work provides a practical realization of the 60 GHz DMA-based transceiver solution for low-complexity, high-performance, and programmable mmWave communication systems.

## II. System Architecture and Hardware Implementation

### A. Transmitter Implementation

The end-to-end transceiver system architecture is illustrated in Fig. 1. The developed system consists of three primary subsystems: (i) digital baseband processing, (ii) mmWave RF front-end for mmWave frequency translation, and (iii) antenna subsystem, with the Tx and Rx antennas placed 1 m apart.

As shown in Fig. 2(a), at the Tx, we used a USRP B200mini to generate the QPSK video signal and the subsequent RF carrier. The baseband processing was implemented in GNU Radio using the second-generation DVB-S2 (Digital Video Broadcasting) standard to generate a real-time video stream.

A dedicated host PC was connected to the USRP to run the GNU Radio transmitter program. In the DVB-S2 Tx program, a continuous data stream is generated and mapped to QPSK symbols in GNU Radio. Forward error correction with a coding rate of 5/6 was applied to improve link reliability. The resulting complex symbols are pulse-shaped using a root-raised cosine filter to control spectral occupancy and minimize inter-symbol interference. The symbol rate was set to 2 MSymbols/s, at which the video bitstream is encoded. Next, we set the sample rate to 4 MSamples/s, which is the processing rate in GNU Radio and essentially the bandwidth of the carrier IF signal. The IF signal was centered at 1 GHz carrier frequency.

The IF signal was fed into the COTS 60 GHz up-conversion module from Analog Devices, EK1HMC6350 evaluation board integrated with the HMC6300 Tx front-end IC, which translates the signal to the 60 GHz band. The 60 GHz signal was fed to the DMA for OTA transmission. The beam direction of the DMA can be electronically steered by applying coded sequences controlled by an external FPGA for more dynamic scenarios. The detailed design of utilized DMA is presented in [6]. Nevertheless, here for this demonstration, we generated a broadside beam at 60 GHz using a specific binary-coded sequence, with 10 dBi gain, and tested the link in a single direction. Nonetheless, DMA can offer dynamic beam steerability through FPGA coding and multi-directional link can be established.

### B. Receiver Implementation

At the Rx side, as shown in Fig. 2(b), the transmitted signal was captured using a V-band horn antenna. It was subsequently down-converted using a corresponding 60 GHz HMC6301 Rx module. The recovered IF signal was then fed into the X300 USRP, where the implemented receiver program performed digital demodulation and real-time video reconstruction. For video reconstruction, we used SDRangel software and implemented a DATV demodulator for video reception. SDRangel is an open-source SDR environment that provides versatile RF features, including real-time spectrum, overlapped spectrum, 2D and 3D spectrograms, a time-domain oscilloscope, statistical values of signal-to-noise ratio (SNR), carrier-to-noise ratio (CNR), channel power, occupied bandwidth, and digital TV signal reconstruction.

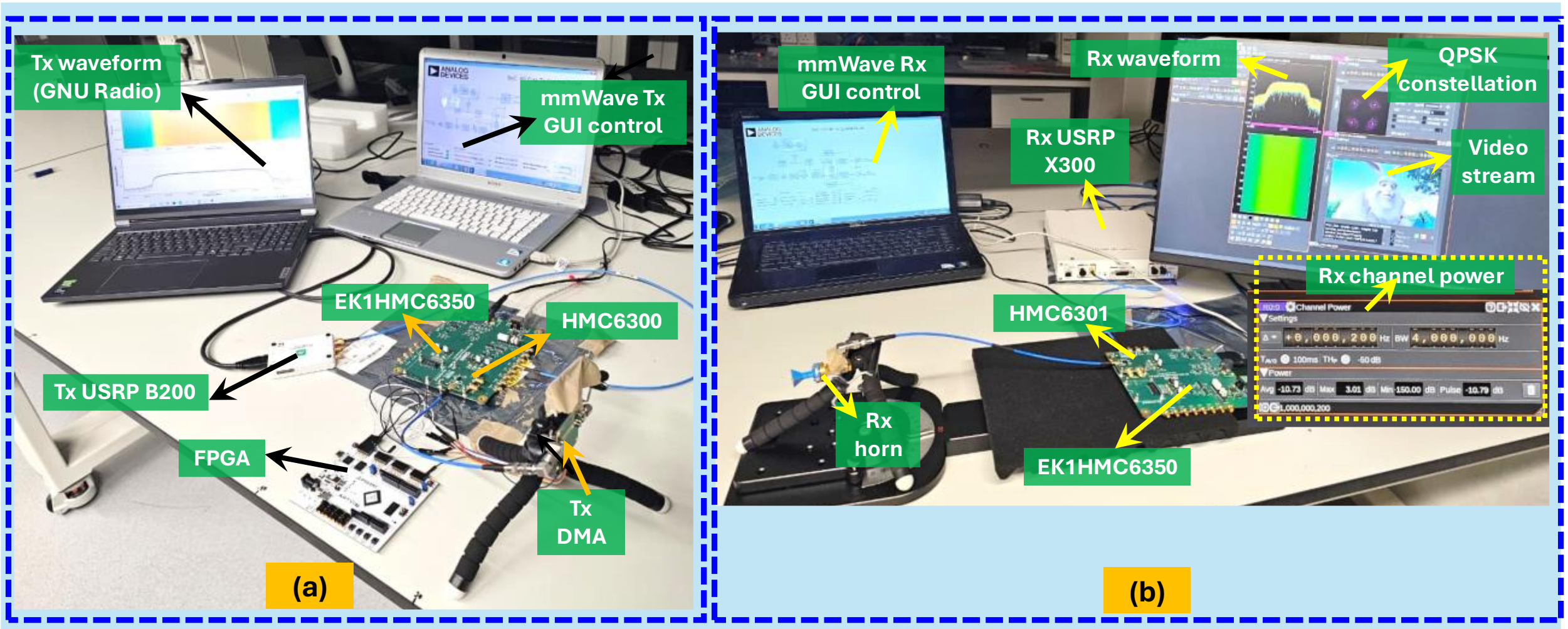

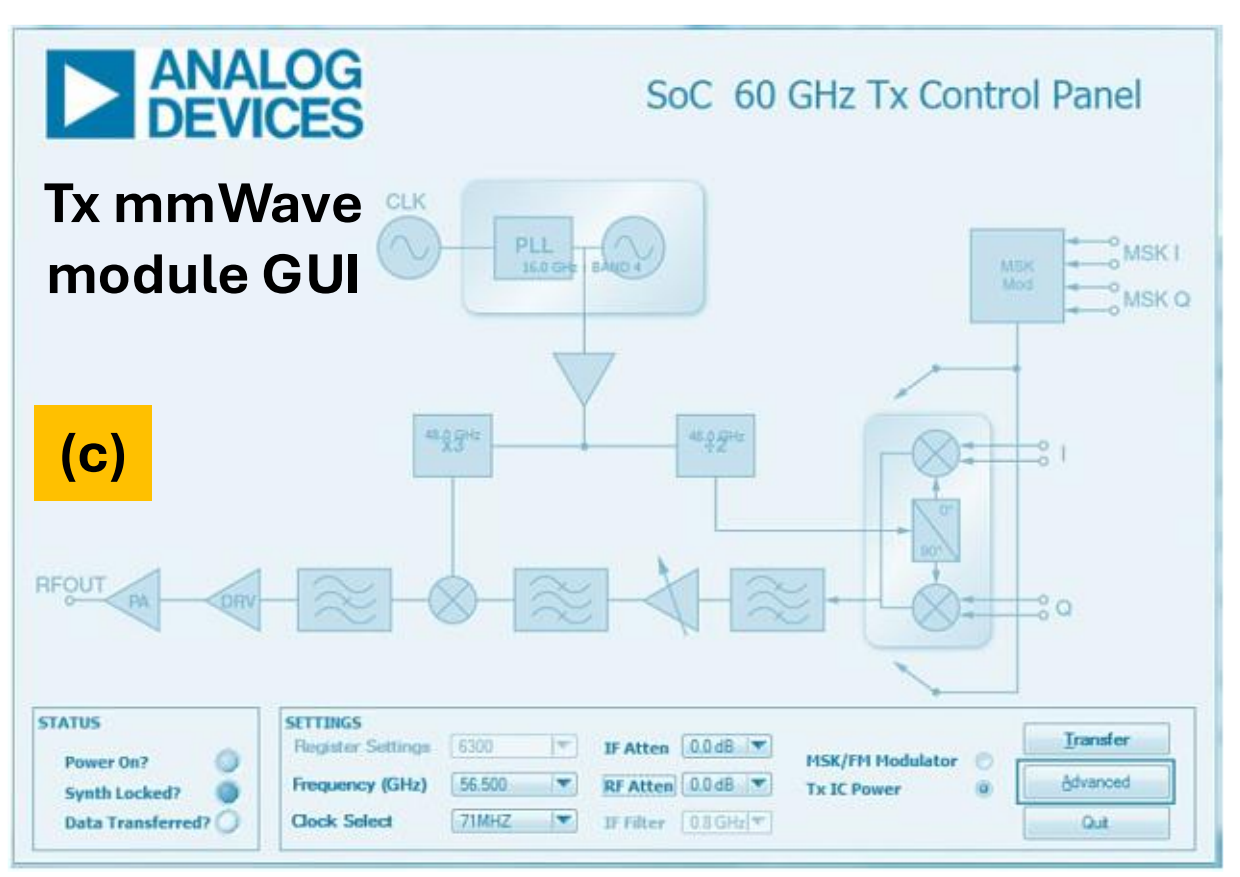

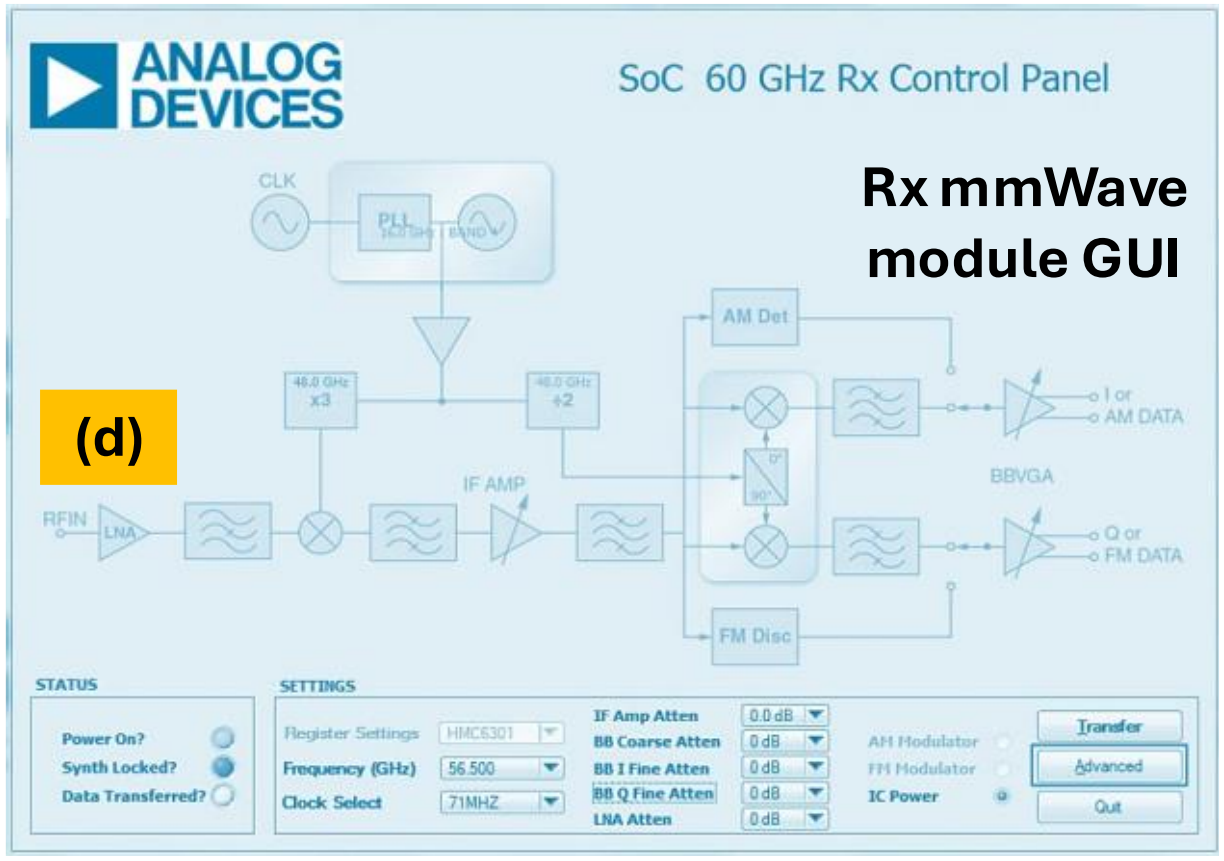

Fig. 2. (a) Experimental transmitter setup, including GNU Radio baseband processing and the mmWave up-conversion module feeding the DMA. (b) Experimental receiver setup with down-conversion module.(c) 60 GHz Tx Control Panel Window for HMC6300 Tx GUI. (d) HMC6301 Rx GUI.

### *C. RF Frontend and Up/Down Conversion*

The mmWave RF front-end plays a critical role in allowing mmWave transmission. Due to hardware constraints of utilized USRP devices up to a maximum of 6 GHz, direct generation of 60 GHz signals is not feasible. Therefore, a heterodyne architecture is employed. We used the EK1HMC6350 evaluation board at both the Tx and Rx sides, which offers a half-duplex, 60 GHz mmWave link using standard baseband analog interfaces. It includes the Tx and Rx module assemblies, USB control, baseband interfaces, and MMPX-to-1.85 mm RF connections for connectorized 60 GHz antennas. A dedicated USB interface and graphical user interface (GUI) software are available to control the settings of the Tx module (HMC6300) (Fig. 2(c)) and Rx module (HMC6301) (Fig. 2(d)). We used dedicated PCs to control Tx and Rx mmWave modules through the GUI. It is instructive to mention here that due to legacy firmware constraints of these evaluation boards, earlier operating systems (e.g., Windows 7) are required for proper operation of the control GUI. The HMC6300/6301 modules have an integrated synthesizer that provides tuning in 250 MHz, 500 MHz, or 540 MHz steps with excellent phase noise. The supported frequency range is between 57 GHz and 64 GHz.

***Up-Conversion****:* For up-conversion, an integrated frequency synthesizer in HMC6300 IC creates a low phase noise LO signal between 16.3 GHz and 18.3 GHz. For any external input IF signal (such as 1 GHz signal from the Tx USRP in this case), the signal is quadrature modulated onto an 8 GHz to 9.1 GHz sliding IF using the synthesized LO divided by two (*IF = LO/2*). The IF signal is then filtered and amplified with 14 dB of variable gain, then mixed with three times the LO frequency to upconvert to an RF frequency between 57 GHz and 64 GHz. In other words, for up-conversion, *RF = IF + LO; therefore, as per the above-described* heterodyne architecture, *RF = LO/2 + 3(LO).* From the GUI, we can only control and lock the RF frequency.

***Down-Conversion****:* At the Rx side, when a 57 GHz to 64 GHz signal enters the HMC6301 module, its integrated LNA provides 20 dB of variable gain. The internally generated LO (16.3 GHz to 18.3 GHz) is multiplied by three and mixed with the LNA output (incoming RF) to down-convert to an 8.14 GHz to 9.1 GHz sliding IF. An integrated notch filter removes the image frequency from 40 GHz to 46 GHz. The IF signal is filtered and amplified with a variable gain of 14 dB, and fed into a quadrature demodulator using the *LO/2* to down-convert to baseband.

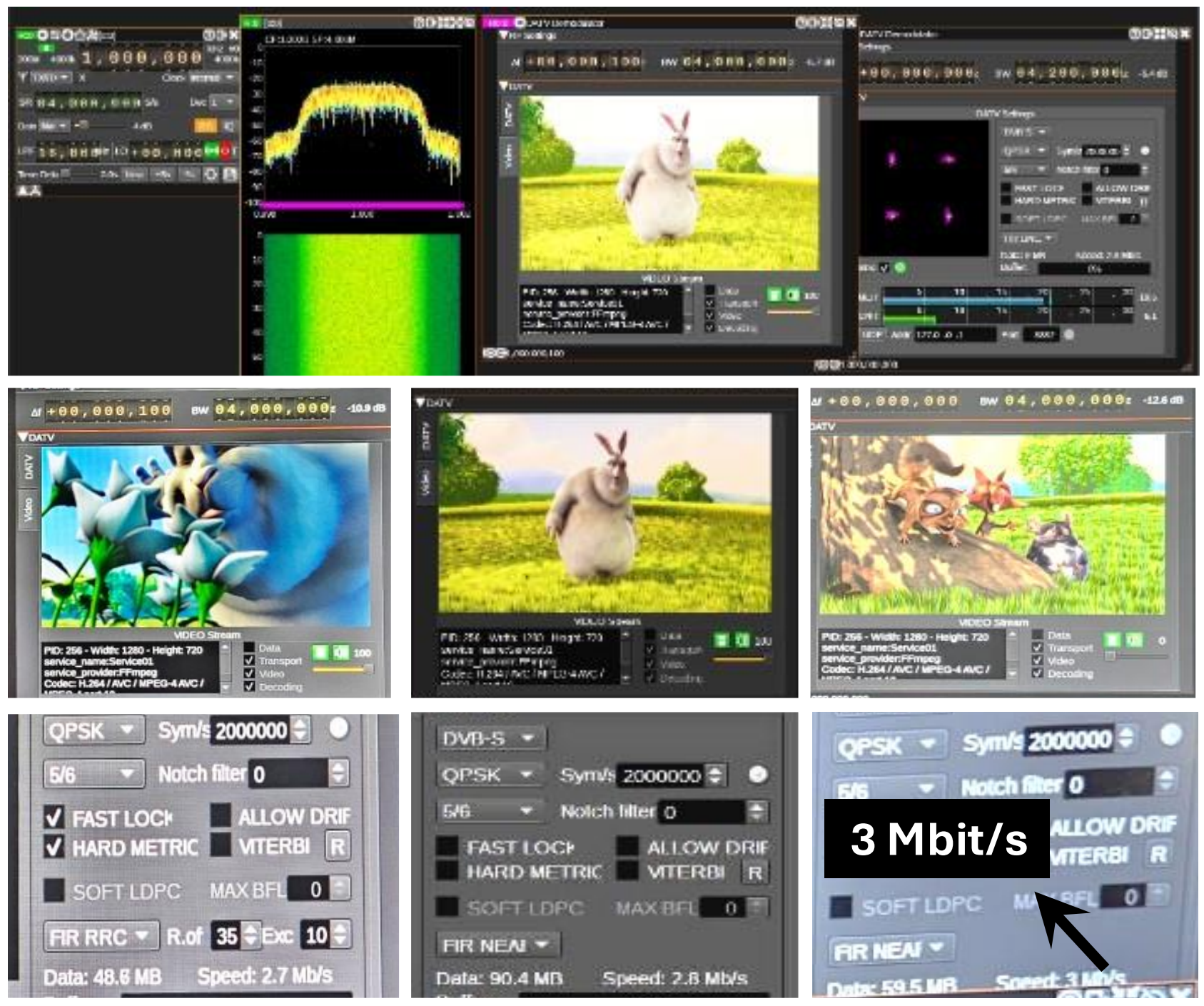

Fig. 3. Received signal profile and experimental real-time snapshots of the received HD (1280 × 720) video stream, demonstrating an average data rate of 3 Mbit/s.

## III. Theoretical Data Rate Prediction and Experimentally Measured Performance

### A. Theoretical Data Rate Calculation

For the implemented communication setup, we can determine the theoretical maximum raw PHY data rate by the utilized symbol rate, modulation order, and coding rate as: $Data\ rate = \frac{symbol}{sec} \times \frac{bits}{symbol} \times code\ rate$. For QPSK, which carries 2 bits per symbol (four constellation points), a coding rate of 5/6 and a symbol rate of 2 MSymbols/s yield a theoretical maximum data rate of 3.33 Mbit/s.

Note that the motivation for operating at 60 GHz lies beyond raw capacity. Operating at 60 GHz enables highly directive transmission, reduced interference, and improved spatial reuse, which are not achievable with low-frequency sub-6 GHz omnidirectional systems. Furthermore, the large available bandwidth at mmWave frequencies provides significant scalability toward higher data rates possible with increased symbol rates and advanced modulation schemes.

### B. Experimental Results

The Tx and Rx nodes were separated by 1 m under LoS conditions. Due to the narrow beamwidth and high directivity of DMA, precise alignment between the Tx and Rx antennas is critical. Fine alignment was achieved through iterative adjustments using a laser alignment tool to maximize received signal quality. The receiver parameters, including carrier frequency (1 GHz), symbol rate, and coding rate, were configured in the SDRangel software to match the transmitter settings.

The transmission was initiated through GNU Radio, and the received signal quality was monitored in real time using the SDRangel spectrum analyzer tool, which provides metrics such as CNR and constellation visualization. Subsequent fine-tuning of antenna orientation and gain settings within the GNU Radio and USRP chain was performed to optimize the received constellation and waveform integrity.

A continuous HD video stream was successfully transmitted and recovered at the receiver, as illustrated in Fig. 3. The received QPSK signal exhibited a measured CNR of approximately 6 dB and a modulation error ratio (MER) of approximately 12 dB, corresponding to an estimated RMS error vector magnitude (EVM) of approximately 25%. The measured average data rate was approximately 3 Mbit/s under these test conditions, corresponding to approximately 90% of the theoretically calculated PHY data rate of 3.33 Mbit/s. The observed ~10% reduction is attributed to a combination of practical factors, including DVB-S2 framing overhead, SDR processing latency, buffering, and residual link impairments such as minor misalignment and hardware non-idealities.

## IV. Conclusion

This work demonstrates a practical 60 GHz DMA-enabled wireless communication system capable of real-time HD video transmission. An end-to-end experimental platform, integrating USRPs with mmWave front-end modules, is successfully realized. A 720p (1280 × 720) video stream is reliably transmitted between the transmitter and receiver, achieving a data rate of approximately 3 Mbit/s with stable constellation performance. The results demonstrate the feasibility of a programmable DMA as a viable antenna architecture for integrated communication systems. Importantly, this work establishes a clear pathway toward scalable mmWave links, with potential extensions to higher-order modulation schemes, wider bandwidths, and multi-beam operation for multi-user scenarios. Furthermore, the proposed architecture is scalable across frequency bands, supporting applicability in high-throughput wireless systems.